\documentclass[numberedappendix,apj,twocolumn,letter]{emulateapj}

\usepackage{amssymb}
\usepackage{amsmath}
\usepackage{lscape}
\newcommand{\be}{\begin{equation}}
\newcommand{\ee}{\end{equation}}

\shorttitle{The BORG survey} 
\shortauthors{Trenti et al.}

\begin{document}


\title{The \emph{Brightest Of Reionizing Galaxies} Survey: Design and Preliminary Results \footnote{Based on observations made with the NASA/ESA Hubble
Space Telescope, which is operated by the Association of Universities for Research in Astronomy, Inc., under NASA contract NAS
5-26555. These observations are associated with Programs $11700, 11702$}}

\author{M. Trenti} \affil{University of Colorado, Center for Astrophysics and Space Astronomy, 389-UCB, Boulder, CO 80309 USA} \email{trenti@colorado.edu} 
\and
\author{L.~D. Bradley}\affil{Space Telescope Science Institute, 3700 San Martin Drive Baltimore MD 21218 USA}
\and \author{M. Stiavelli} \affil{Space Telescope Science Institute, 3700 San Martin Drive Baltimore MD 21218 USA}
\and \author{P. Oesch} \affil{Institute of Astronomy, ETH Zurich, CH-8093 Zurich, Switzerland}
\and \author{T.~Treu}\affil{Department of Physics, University of California, Santa Barbara, CA 93106-9530, USA}
\and \author{R.~J. Bouwens} \affil{Astronomy Department, University of California, Santa Cruz, CA 95064, USA ; Leiden Observatory, University of Leiden, Postbus 9513, 2300 RA Leiden, Netherlands}
\and \author{J.~M. Shull}\affil{CASA, Department of Astrophysics and Planetary Science, University of Colorado, 389-UCB, Boulder, CO 80309 USA}
\and \author{J.~W. MacKenty}\affil{Space Telescope Science Institute, 3700 San Martin Drive Baltimore MD 21218 USA}
\and \author{C.~M. Carollo} \affil{Institute of Astronomy, ETH Zurich, CH-8093 Zurich, Switzerland}
\and \author{G.~D. Illingworth} \affil{Astronomy Department, University of California, Santa Cruz, CA 95064, USA}



\begin{abstract}
 
  We present the first results on the search for very bright ($M_{AB}
  \approx -21$) galaxies at redshift $z\sim 8$ from the Brightest of
  Reionizing Galaxies (BoRG) survey. BoRG is a Hubble Space Telescope
  Wide Field Camera 3 pure-parallel survey that is obtaining images on
  random lines of sight at high Galactic latitudes in four filters
  (F606W, F098M, F125W, F160W), with integration times optimized to
  identify galaxies at $z\gtrsim 7.5$ as $F098M$-dropouts. We discuss
  here results from a search area of approximately $130$ arcmin$^2$
  over 23 BoRG fields, complemented by six other pure-parallel WFC3
  fields with similar filters. This new search area is more than two
  times wider than previous WFC3 observations at $z\sim 8$.  We
  identify four F098M-dropout candidates with high statistical
  confidence (detected at greater than $8\sigma$ confidence in F125W). These sources
  are among the brightest candidates currently known at $z\sim 8$ and
  approximately ten times brighter than the $z=8.56$ galaxy
  UDFy-38135539. They thus represent ideal targets for spectroscopic
  followup observations and could potentially lead to a redshift
  record, as our color selection includes objects up to $z\sim
  9$. However, the expected contamination rate of our sample is about
  $30\%$ higher than typical searches for dropout galaxies in legacy
  fields, such as the GOODS and HUDF, where deeper data and additional
  optical filters are available to reject contaminants.

\end{abstract}

\keywords{galaxies: high-redshift --- galaxies: evolution}

\section{Introduction}

The installation of Wide Field Camera 3 (WFC3) on the \emph{Hubble Space
Telescope} opened new possibilities for discovery of $z>7$
galaxies. Observations on the GOODS and HUDF fields have
increased the sample of $z\gtrsim 7$ candidates to $N>100$
\citep{oesch09_zdrop,bouwens10c,mclure09,finkelstein10,wilkins10}.
Legacy multi-wavelength coverage on these fields and the improved
spatial resolution of WFC3 enabled preliminary characterization of the
properties of these sources in terms of stellar mass, stellar
populations and size \citep{oesch09_size,bouwens09_slope,labbe09a}.

However, the current search area, while containing deep and ultradeep
data, is modest (approximately $60$ arcmin$^2$) and located within or around a
single GOODS field. This provides significant uncertainty in the
number density of $z\gtrsim 7$ galaxies owing to small-number
statistics and cosmic variance \citep{bouwens10c}. This is especially
severe at the bright end of the luminosity function (LF), where
sources are most clustered and least abundant (e.g., see
\citealt{trenti08}). These WFC3 observations suggest that the galaxy
LF evolves sharply from $z\sim 6$ to $z\sim 8$, particularly at the
bright end \citep{bouwens10c}. Such trend is consistent with the
underlying evolution of the dark matter halo mass function, which
predicts well the LF evolution \citep{trenti10b}, but
stronger observational constraints on $M_*$ are needed.

Reducing uncertainty on the number density of bright $z\sim 8$ sources
also benefits the determination of the LF faint-end slope $\alpha$.
Fits to a Schechter (1976) LF, $\phi(L)=\phi_* (L/L_*)^{\alpha}
\exp{(-L/L_*)}$, have a strong degeneracy between characteristic
luminosity $M_* = -2.5 \log_{10}(L_*)$ and $\alpha$
\citep{bouwens07,trenti08}. Measuring $\alpha$ is fundamental to
assess whether galaxies emit enough photons to reionize the Universe
\citep{bunker04,chary08,henry09,trenti10b,robertson10}. Ground-based programs \citep{ouchi09,castellano10}
place useful constraints on $M_*$ at slightly lower redshift
($z \lesssim 7$), but the impact of large-scale structure remains a
concern, because all these searches are within legacy fields.

To complement the existing campaigns aimed at searching for
$z\gtrsim7.5$ galaxies, we introduce here the HST BoRG (Brightest of
Reionizing Galaxies) survey. BoRG is based on pure-parallel
observations with HST-WFC3 while the telescope is pointing to a
primary spectroscopic target (typically a foreground high-$z$ QSO).
Because lines of sight are independent and well separated on the sky,
cosmic variance is negligible. In contrast, cosmic variance introduces
approximately $25\%$ uncertainty, in addition to Poisson noise, in the
number counts of $z\gtrsim 6$ galaxies for both the GOODS and the HUDF
surveys. A survey with independent lines of sight 
provides an unbiased characterization of the LF bright-end with errors
equivalent to those of a continuous survey of about twice its area
\citep{trenti08}.

The preliminary BoRG dataset discussed here contains 29 lines of sight
for a total of approximately $130~\mathrm{arcmin^2}$, more than twice
the area of the HUDF and GOODS-ERS observations in
\citet{bouwens10c}. This wide area allows us to identify galaxies that
are good candidates for follow-up spectroscopic observations; all
$z\sim 8$ galaxies in BoRG are significantly brighter than
UDFy-38135539 for which \citet{lehnert10} reported detection of
Ly$\alpha$ emission at $z=8.56$.
 
Section~\ref{sec:survey} of this paper introduces the BoRG survey.
Data reduction is discussed in Section~\ref{sec:data}.
Section~\ref{sec:selection} presents our selection strategy along with
estimates of contamination. Preliminary results are
presented in Section~\ref{sec:results} and compared with an independent
analysis by \citet{yan10}. Section~\ref{sec:conclusion} summarizes and
concludes. We adopt a standard WMAP7 cosmology \citep{wmap7} and the
AB magnitude scale \citep{oke}.

\section{Survey Design}\label{sec:survey}

The BoRG survey is designed to identify bright ($m_{F125W} \lesssim 27
$) high-redshift galaxies from their broad-band colors using the
Lyman-Break technique \citep{steidel96}. The primary aim of the survey
is to select $z\gtrsim 7.5$ galaxies as F098M dropouts. Two near-IR
filters (F125W and F160W) are used for source detection. One optical
filter (F606W) is used to control the primary source of contamination
from lower redshift $z\sim 1.5$ interlopers (see
Section~\ref{sec:selection}). As we detail below, the survey was
designed to minimize the probability that artifacts and low-redshift
interlopers may pass our selection criteria.

Parallel opportunities of program GO/PAR 11700 are at least three
orbits in length (mostly 3-5 orbits). Each individual
visit has a particular duration determined by the details of its
primary program (see Table~\ref{tab:fields}). The exposure time
between filters has been allocated by keeping the relative depths
approximately constant, within the constraints imposed by the primary
program. In calculating the relative exposure times, we also took into
account the effect of Galactic reddening. We used the primary line of
sight as a proxy for the extinction expected in WFC3
images. From the reddening map of \citet{schlegel98} and
extinction law of \citet{card89}, we derived the extinction
in each band.

Dithering in pure-parallel observations is determined by the primary
program, so it is usually absent because primary observations are
spectroscopic. This introduces some challenges in the data analysis,
especially with respect to systematic errors introduced by the
detector, namely hot pixels and detector persistence in the IR
channel, which arises following observations of targets that saturate
the detector \citep{dressel}. The latter is of particular concern
because of the possibility of introducing an artificial coherent
signal into the detection band(s) for F098M-dropouts, thereby leading
to false candidates. To minimize the impact of persistence, we ensured
that every observation of Program 11700 in either F125W or F160W is
preceded in the same orbit by a comparably long F098M exposure. When
possible, F160W observations follow F125W. As detector persistence
decays over time (with approximate power-law behavior), any saturated
target observed in a previous visit affects most the initial part of
the pure-parallel orbit, which is the exposure in the dropout
filter. With this strategy, persistence features are guaranteed not to
contaminate the dropout selection. To ensure good sampling of the IR
exposures, we opted for reading every 100 s (SPARS100). While the
majority of cosmic rays are rejected by the calibration pipeline,
owing to the multiple non-destructive readouts of WFC3, a small
fraction may survive in the calibrated image. We thus split the total
integration in each IR filter into at least two individual
exposures. F606W exposures are split in $N_s\geqslant 3$ sequences
(each $500-900$ s) for cosmic-ray rejection, except for shallower
fields, where $N_s=2$ if the total F606W integration is less than
$1000~\mathrm{s}$. Our design choices are aimed at maximizing the data
quality, although a small price is paid in the term of signal-to-noise
ratio. For example, our strategy to ``shield'' F125W and F160W
observations from persistence by means of a F098M exposure carries
some overhead because filter rotation happens during the observation
window.

We also consider a small number of fields from another pure-parallel
program with the same IR filters but with F600LP instead of F606W
(GO/PAR 11702, PI Yan). Images in program GO/PAR 11702 are not
characterized by the optimization described above; for a given
pointing, F098M exposures tend to be in different orbits than the
redder IR filters. In addition, some IR filters only have a single
exposure and both SPARS100 and SPARS200 sampling is used. Overall,
this makes the additional dataset potentially more vulnerable to
spurious sources.

\section{Data Reduction and Catalog Construction}\label{sec:data}

The images were reduced using standard techniques. For the WFC3/IR
data, we recalibrated the raw data using calwfc3 using the most
up-to-date reference files and our own custom flat fields generated by
median stacking of publicly available WFC3/IR data in F098M, F125W, and
F160W. We used SExtractor \citep{bertin96} to background subtract
the FLT files prior to running multidrizzle \citep{drizzle}. The background levels
were stored in the headers of each FLT file and are subsequently used
by multidrizzle for cosmic-ray rejection. The individual exposures
were aligned and drizzled on a common $0.08$ arcsec/pixel scale using
multidrizzle.

To identify F098M-dropout sources we first constructed a preliminary
catalog with SExtractor, then we checked the source S/N reported by
SExtractor and normalized the input rms maps if needed. Finally
we reran SExtractor to obtain the final source catalog. Below we
describe these steps.

In each field, we identified sources from the F125W image using
SExtractor in dual image mode. We required at least 9 contiguous
pixels with $S/N \geqslant 0.7$ for the preliminary catalog.

Multidrizzle introduces correlated noise in the images
\citep{casertano00}.
To derive realistic errors, one needs to rescale the rms map by
measuring the noise in areas of size comparable to observed
galaxies. Therefore, we selected 400 random pointings at distance
$d>0.4\mathrm{"}$ from detected sources. There we performed circular
aperture photometry (radius $r=0.32\mathrm{"}$) with SExtractor in dual
image mode. We used a synthetic detection image with artificial
sources at the location of the random pointings and the actual images
as photometry frames. We normalized the rms map of each filter
requiring that, for these sky apertures, the median of the nominal
error reported by SExtractor (FLUXERR\_APER) is equal to the rms of
the measured flux (FLUX\_APER). This results in an average
multiplication of the rms maps by $1.5$ for F606W, by $1.15$ for F098M
and by $1.1$ for F125W and F160W.

After rescaling the rms maps, we reran SExtractor to create a final
version of the catalogs. To include a source, we required detection
with $S/N>8$ in F125W and $S/N>3$ in F160W for ISOMAG fluxes. Colors
were computed from ISOMAG measurements.  Total magnitudes were defined
as AUTOMAG.

We derived median $5\sigma$ sensitivities in a circular aperture with
radius $r=0.32\mathrm{"}$ (median exposures times also listed) of
$m_{F606W}= 26.9$ ($t_{exp}=2647$ s), $m_{F600LP}=26.4$
($t_{exp}=2334$ s), $m_{F098M}= 26.8$ ($t_{exp}=4515$ s), $m_{F125W} =
26.7$ ($t_{exp}=2205$ s), $m_{F160W}= 26.3$ ($t_{exp}=1405$
s). Table~\ref{tab:fields} reports the individual field sensitivities.
We used photometric zero points $26.08$, $25.85$, $25.68$, $26.25$,
$25.96$ respectively for F606W, F600LP, F098M, F125W, F160W
\citep{dressel}.

\section{Selection of $z \sim 8$ Galaxy Candidates}\label{sec:selection}

Candidate galaxies at $z \gtrsim 7.5$ are selected using broad-band
colors analogous to other $z \gtrsim 6$ galaxy surveys
\citep{oesch09_zdrop,bouwens09_ydrop}. In short, we search for objects
that have a strong break in the filter corresponding to the redshifted
Ly$\alpha$ absorption at $z\gtrsim 7.5$. We use F098M as the dropout
filter, requiring:
\begin{equation}
m_{F098M}-m_{F125W} \geq 1.75.
\end{equation}
If $S/N<1$ in F098M, we replace the measured flux with its $1\sigma$
limit. Furthermore, to remove lower redshift red and/or obscured
contaminants, we require that $m_{F125W}-m_{F160W}$ is moderately red at most:
\begin{equation}\label{eq:J-H}
m_{F125W}-m_{F160W} < 0.02 + 0.15 \times (m_{F098M}-m_{F125W}-1.75).
\end{equation}
The third condition we impose is a conservative non-detection at
$1.5\sigma$ in the optical band available (F606W or F600LP).

\begin{figure} 
  \plottwo{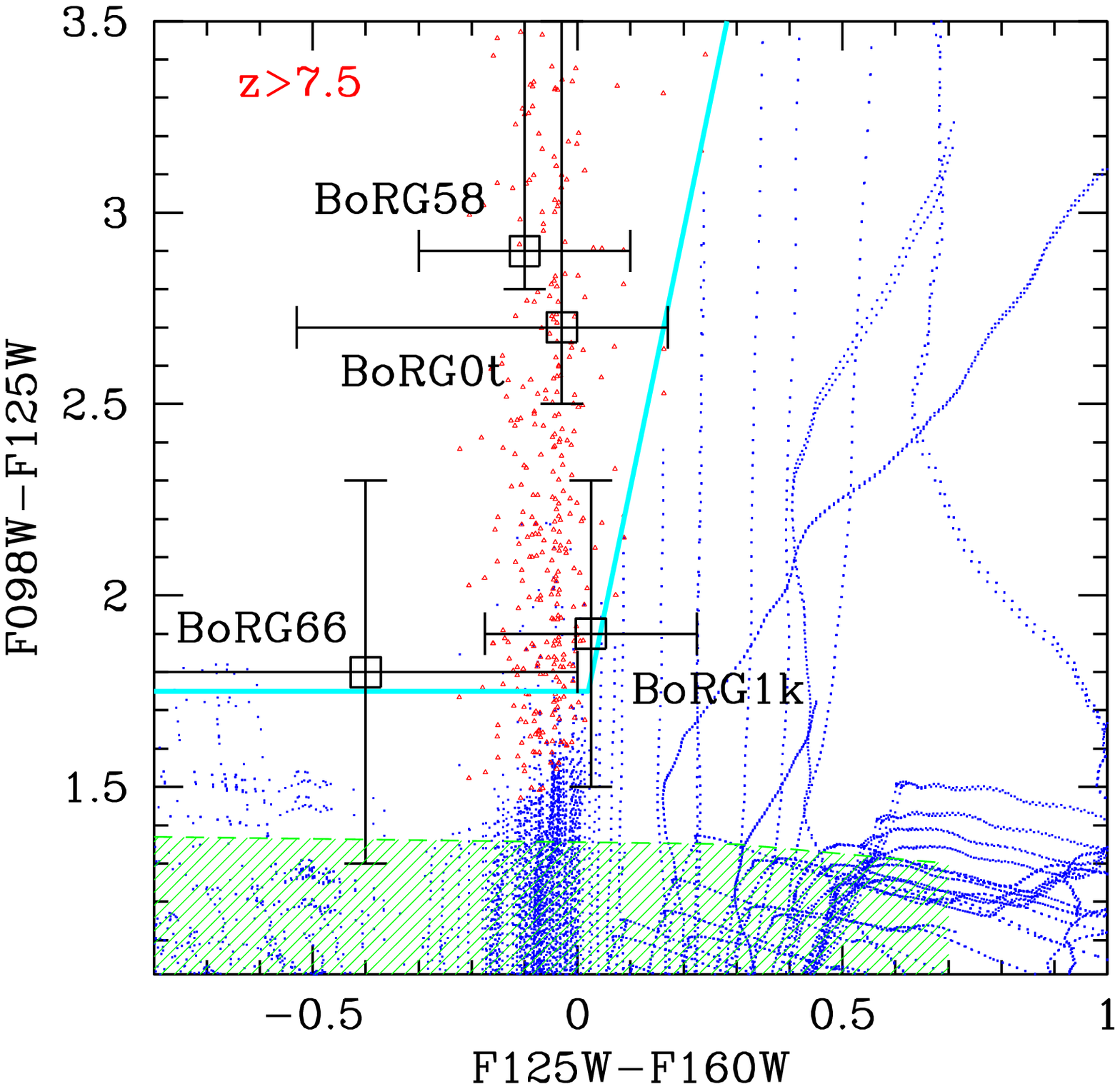}{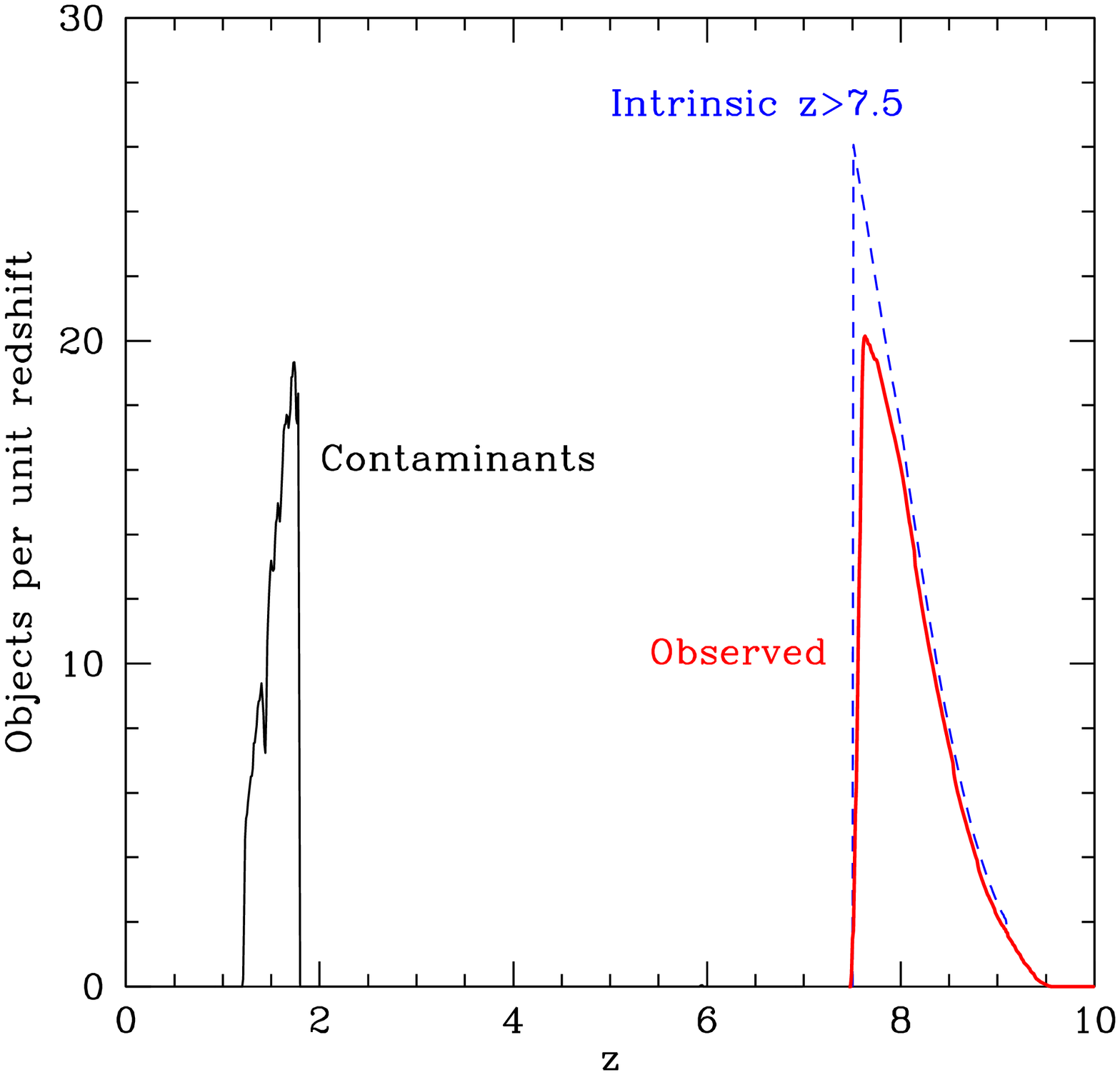} \caption{Left: F098M-dropouts
    color-color selection. Black squares indicate our four $z>7.5$
    candidates (with $1\sigma$ error bars). 
Cyan lines denote selection window. Blue dots are 
    simulated low-redshift interlopers.  Red triangles $z>7.5$
    galaxies. L, T dwarf stars from \citet{knapp04} occupy green shaded area (see \citealt{bouwens10c}).  Right: Expected redshift distribution for candidates within the selection window  (red: $z>7.5$ galaxies; black: low-$z$ interlopers).    Blue-dotted line shows redshift distribution for
    luminosity-limited selection of $z>7.5$ galaxies (the color-color selection rejects some sources).
  }\label{fig:color_sel}\label{fig:zdis}
\end{figure}

These conditions have been chosen by optimizing the selection
efficiency of genuine $z\gtrsim 7$ galaxies while minimizing
contamination from low-redshift galaxies and cool stars. Our IR
color-color selection window is shown in Fig.~\ref{fig:color_sel},
along with typical colors for possible galaxy contaminant sources. The
figure is based on a library of 10 million galaxy spectral energy
distributions constructed using different star-formation histories,
metallicities, and dust content (see
\citealt{oesch07}). Fig.~\ref{fig:zdis} also shows the expected
redshift distribution of Lyman Break galaxies entering our selection
window.

The availability of high-quality deep optical data appears to be the
limiting factor in the rejection of low-redshift galaxy
contaminants. In fact, passively evolving galaxies at $z\sim 1.5$ can
contaminate the F098M-dropout selection when their
$4000~\mathrm{\AA}$/Balmer break is misidentified as Ly$\alpha$ break
if the data are not sufficiently deep to detect these sources in the
optical bands \citep[see also][]{henry09,capak09}. We estimate
approximately $30\%$ contamination using the GOODS-ERS data (Program
11359) which include F098M imaging. We estimate this number as
follows: from the \citet{bouwens10c} data reduction we first identify
F098M-dropouts with $m_{F125W} \leq 27$ using the selection discussed
above, but considering a version of the GOODS $F606W$ image degraded
to a $5\sigma$ limit $m_{F606W}=27.2$ to match the relative F125W
vs. F606W BoRG depth. We then check for contaminants by rejecting
F098M-dropouts with $S/N>2$ in either $B$, $V$, or $i$ (at their full
depth). Approximately $30\%$ contamination is in good agreement with
the estimate based on the application of the color selection to our
library of SED models (Figure~\ref{fig:zdis}).

Cool stars are another possible source of contamination. Our survey
area is large and probes lines of sight at different Galactic
latitudes. Thus, we cannot argue that they are unlikely to be present
based on their rarity as in the HUDF \citep{bouwens09_ydrop}.
However, based on the brown dwarf spectra of \citet{reid01} and on the
colors for L and T dwarfs measured by \citet{knapp04}, the use of
F098M as a dropout filter is efficient at rejecting brown dwarf stars;
the expected colors of these contaminants are well separated from our
selection window (see Figure~\ref{fig:color_sel}). In addition, owing
to the high-resolution of WFC3, it is expected that all the brightest
galaxies at $z\gtrsim 7$ will be extended sources
\citep{oesch09_size}, providing a further diagnostic to identify
Galactic stellar contaminants. Our bright dropouts ($m_{F125W}\leq
26$) have low values of the SExtractor Stellarity parameter (see
Table~\ref{tab:candidates}), hence they are consistent with being
resolved.  Overall, we do not expect this source of contamination to
be significant.

\section{Sample of $z\sim 8 $ candidates}\label{sec:results}

\begin{figure} 
  \plotone{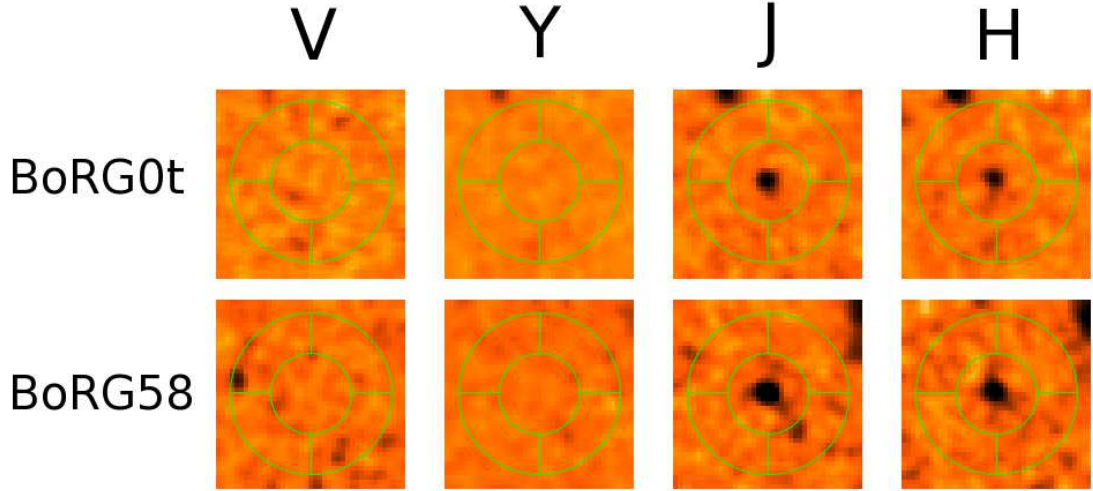} \caption{Region ($3.2\mathrm{"}\times
    3.2\mathrm{"}$) surrounding the most robust BoRG
    F098M-dropout candidates (left to right: F606W, F098M, F125W,
    F160W).}  \label{fig:stamps} \end{figure}

\begin{figure} 
\plotone{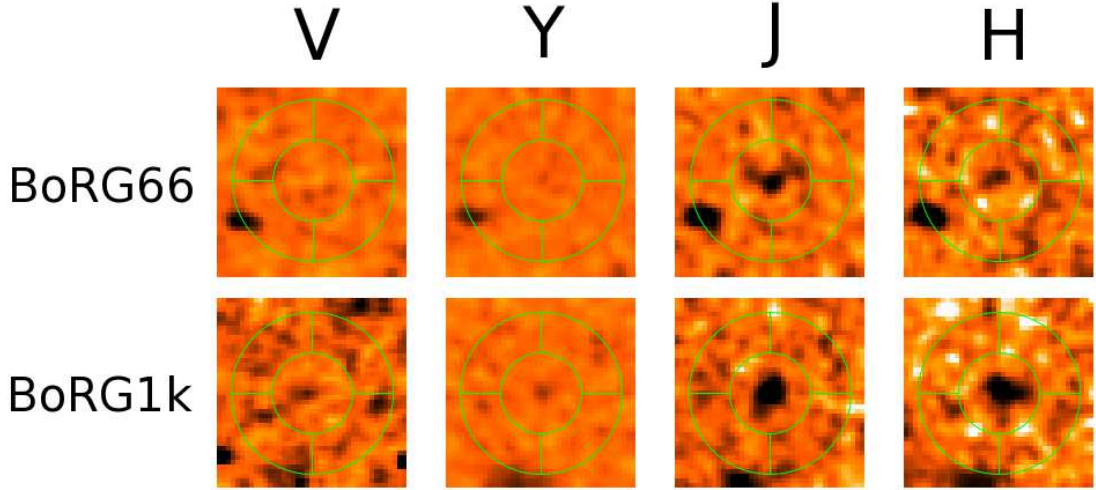} \caption{As in Fig.~\ref{fig:stamps}, but showing the least robust BoRG dropouts.}  \label{fig:stamps2} \end{figure}

Four objects satisfy our dropout selection, all from BoRG fields.
Their photometry is reported in Table~\ref{tab:candidates}.
Figures~\ref{fig:stamps}-\ref{fig:stamps2} show the candidates'
images. Here we discuss each candidate, critically assessing its
likelihood to be at $z\gtrsim 7.5$, starting with the least
robust. Three of these four objects have been identified as
F098M-dropouts by an independent analysis of our data
\citep{yan10}. All fields with candidates contain at least two
exposure frames per filter (taken in different orbits).  We verified
that candidates are detected in each individual F125W and F160W frame
at S/N ratio consistent with scaling from the total to the
individual exposure time.

\subsection{BoRG66\_1741-1157}

This object is detected in $F125W$ with $S/N=8.7$, but only marginally
in $F160W$ ($S/N=3.3$). Hence, it has the bluest $J-H$ color of the
sample. With $Y-J = 1.9\pm 0.6$, it lies at the edge of our dropout
selection window, and its membership in the sample of $z\gtrsim 7.5$
candidates could be the result of photometric scatter. Its blue color
could be due to contribution from strong Ly$\alpha$ emission in
F125W. This object is not in the \citet{yan10} sample.

\subsection{BoRG1k\_0847-0733}

This candidate also lies at the edge of the selection window, and its
inclusion in the sample could be due to photometric scatter in either
$Y-H$ or $J-H$ color (with $\sim 60\%$ probability if errors are
symmetric). There are hints of flux at optical wavelength:
$(S/N)_{F606W} = 1.3$ ($S/N\geq 1.3$ has probability $p\leq 6.6\%$ for
Gaussian noise). The F606W exposure is shallow
(Table~\ref{tab:fields}): $t=1260$ s ($5\sigma$ mag limit $26.3$). The
dropout filter has a marginal detection ($S/N_{F098M}=2.4$). Because
of these multiple bits of circumstantial evidence, we consider the
object a low-probability $z\gtrsim 7.5$ candidate, more likely to be a
passive $z\sim 1.5$ galaxy.

\subsection{BoRG0t\_0958-0641}

BoRG0t\_0958-0641 is the faintest candidate in the sample, but has
$(S/N)_{F125W}>8$ and $(S/N)_{F160}=6.8$ because of the significant
exposure time in this field (Table~\ref{tab:fields}). Data from
programs 11700 and 11702 cover the region at slightly different
orientations, providing some dithering. The candidate is well within
the color-color selection window for $z\gtrsim 7.5$ galaxies
(Figure~\ref{fig:color_sel}). \citet{yan10} do not consider this
object a strong $z>7.5$ candidate because they claim variability on
the three-day timescale of the observations (see their Figure 4). We
performed aperture photometry ($r=0.32\mathrm{"}$) at the source
location for the three epochs, and find no evidence of variability in
F125W (measuring $m=26.7, 26.8, 26.8$ with typical $1\sigma$ sky
uncertainty of approximately $0.35$ mag in each frame). We see
evidence of variability in F160W at about $2\sigma$ confidence:
$m=26.2, 26.9, 27.2$ with $1\sigma$ error of approximately $0.35$
mag. Closer examination highlights potential data quality issues. In
the first epoch, the readouts of the F160W ramp for pixels within the
source jump between readout 8 and 9, indicating a cosmic ray hit. In
the second epoch, there is a hot pixel located within this
source. Because of the stable photometry in F125W, we consider
intrinsic variability unlikely, although further observations would be
useful to clarify the nature of this source.

\subsection{BoRG58\_ 1787-1420}

This is the most robust candidate of the sample, with alll its
properties fully consistent with being a $z \gtrsim 7.5$
galaxy. The measured colors are well within the selection window, even
after taking into account $1\sigma$ errors. The object lies on the $z
\gtrsim 7.5$ galaxy tracks. There is no flux at optical wavelength
($(S/N)_{F606W}<0$).

\section{Discussion and conclusions}\label{sec:conclusion}

In this paper we discuss the preliminary results from the BoRG survey
on the search of bright $z\gtrsim 7.5$ galaxies identified as
F098M-dropouts using HST WFC3 data. By analyzing 29 independent lines
of sight, we identify four dropouts with $(S/N)_{F125W}>8\sigma$. Two
objects lie near the selection window border, and they are possibly
low-redshift interlopers scattered into the selection (but it is
similarly likely that photometric scatter removes objects from the
sample). The remaining candidates satisfy all the expected properties
for $z\gtrsim 7.5$ objects with high confidence.

Detailed discussion on the implications for the evolution of the
galaxy LF are deferred to a future paper (L.~D. Bradley et al., in
preparation). There, we will also attempt to extend the detection of
dropouts to fainter limits and carry out artificial source recovery
simulations to estimate the effective volume of the BoRG survey as a
function of magnitude. Here we consider a magnitude limit $m_{F125W}
\leqslant 26.2$ (equivalent to $M\leqslant -20.9$), where completeness
is close to unity in all regions of the BoRG data not occupied by a
foreground object. To estimate our effective area, we masked all
pixels at distance $d \leqslant 0.4\mathrm{"}$ from a pixel belonging
to the SExtractor segmentation map and counted the remaining pixels,
deriving an effective search area of approximately $97$
arcmin$^2$. Further assuming a pencil-beam geometry with $7.5\leqslant
z\leqslant 8.5$, we derive a comoving volume of $2.3\times 10^5$
Mpc$^3$. From the best fit $z= 8$ LF derived by \citet{bouwens10c}
based on ERS and HUDF data, we expect $N\sim 2.5$ galaxies with
$M\leqslant -20.9$ in our search area. Three candidates at
$m_{F125W}\leqslant 26.2$ are fully consistent with this expectation,
even after taking into account a contamination rate of approximately
$30\%$ from low-$z$ galaxies (Section~\ref{sec:selection}), but
alternative models cannot be strongly ruled out. For constant $\alpha=-2.0$ and $\phi_*=0.38\times 10^{-3}~\mathrm{Mpc^{-3}}$
\citep{bouwens10c}, we derive $M_*=-20.2\pm~0.3$ (68\%
confidence). Quadrupling the BoRG area would allow us to set $\Delta
M_*<0.3$ at $99\%$ confidence. Similar future pure-parallel observations
(e.g. GO/PAR 12286 PI Yan) will contribute toward this goal by
approximately doubling the current search area.


Finally, BoRG58\_1787-1420 represents an ideal candidate for follow-up
spectroscopic investigations. This galaxy is about ten times brighter
than UDFy-38135539 for which \citet{lehnert10} claimed detection of
Ly$\alpha$ emission at $z=8.56$. BoRG58\_1787-1420 could potentially
yield a more secure line identification if the equivalent width is
similar to the $\sim 1900~\mathrm{\AA}$ of UDFy-38135539, or alternatively a
comparable line flux (approximately $6\times 10^{-18}~\mathrm{erg~ s^{-1}
  cm^{-2}}$) for a $200~\mathrm{\AA}$ equivalent width. In addition, from
\citet{trenti10b}, we derive $M_{h} \sim 7 \times 10^{11} M_{\sun}$ as
the host-halo mass for BoRG58\_1787-1420. This galaxy likely lives in
an overdense region of the universe, where the IGM is ionized at early
times, facilitating escape (and detection) of the Ly$\alpha$
radiation.

\acknowledgements 

We thank the referee for useful suggestions. We acknowledge grants
HST-GO-11563 and HST-GO-11700.


\begin{thebibliography}{25}
\expandafter\ifx\csname natexlab\endcsname\relax\def\natexlab#1{#1}\fi

\bibitem[{{Bertin} \& {Arnouts}(1996)}]{bertin96}
{Bertin}, E. \& {Arnouts}, S. 1996, \aaps, 117, 393

\bibitem[{{Bouwens} {et~al.}(2007){Bouwens}, {Illingworth}, {Blakeslee}, \&
  {Franx}}]{bouwens07}
{Bouwens}, R.~J., {Illingworth}, G.~D., {Franx}, M. \& Ford, H. 
  2007, \apj, 670, 928

\bibitem[{{Bouwens} {et~al.}(2010{\natexlab{a}}){Bouwens}, {Illingworth},
  {Oesch}, {Labbe}, {Trenti}, {van Dokkum}, {Franx}, {Stiavelli}, {Carollo},
  {Magee}, \& {Gonzalez}}]{bouwens10c}
{Bouwens}, R.~J., {Illingworth}, G.~D., {Oesch}, P.~A., {Labbe}, I., {Trenti},
  M., {van Dokkum}, P., {Franx}, M., {Stiavelli}, M., {Carollo}, C.~M.,
  {Magee}, D., \& {Gonzalez}, V. 2010{\natexlab{a}}, ArXiv:1006.4360

\bibitem[{{Bouwens} {et~al.}(2010{\natexlab{b}}){Bouwens}, {Illingworth},
  {Oesch}, {Stiavelli}, {van Dokkum}, {Trenti}, {Magee}, {Labb{\'e}}, {Franx},
  {Carollo}, \& {Gonzalez}}]{bouwens09_ydrop}
{Bouwens}, R.~J., {Illingworth}, G.~D., {Oesch}, P.~A., {Stiavelli}, M., {van
  Dokkum}, P., {Trenti}, M., {Magee}, D., {Labb{\'e}}, I., {Franx}, M.,
  {Carollo}, C.~M., \& {Gonzalez}, V. 2010{\natexlab{b}}, \apjl, 709, L133

\bibitem[{{Bouwens} {et~al.}(2010{\natexlab{c}}){Bouwens}, {Illingworth},
  {Oesch}, {Trenti}, {Stiavelli}, {Carollo}, {Franx}, {van Dokkum},
  {Labb{\'e}}, \& {Magee}}]{bouwens09_slope}
{Bouwens}, R.~J., {Illingworth}, G.~D., {Oesch}, P.~A., {Trenti}, M.,
  {Stiavelli}, M., {Carollo}, C.~M., {Franx}, M., {van Dokkum}, P.~G.,
  {Labb{\'e}}, I., \& {Magee}, D. 2010{\natexlab{c}}, \apjl, 708, L69

\bibitem[Bunker et al.(2004)]{bunker04} Bunker, A.~J., Stanway, E.~R.,
  Ellis, R.~S., McMahon, R.~G. 2004, \mnras, 355, 374

\bibitem[{{Capak} {et~al.}(2009){Capak}, {Mobasher}, {Scoville}, {McCracken},
  {Ilbert}, {Salvato}, {Menendez-Delmestre}, {Aussel}, {Carilli}, {Civano},
  {Elvis}, {Giavalisco}, {Jullo}, {Kartaltepe}, {Leauthaud}, {Koekemoer},
  {Kneib}, {LeFloch}, {Sanders}, {Schinnerer}, {Shioya}, {Shopbell},
  {Tanaguchi}, {Thompson}, \& {Willott}}]{capak09}
{Capak}, P., {Mobasher}, B., {Scoville}, N.~Z., {McCracken}, H., {Ilbert}, O.,
  {Salvato}, M., {Menendez-Delmestre}, K., {Aussel}, H., {Carilli}, C.,
  {Civano}, F., {Elvis}, M., {Giavalisco}, M., {Jullo}, E., {Kartaltepe}, J.,
  {Leauthaud}, A., {Koekemoer}, A.~M., {Kneib}, J., {LeFloch}, E., {Sanders},
  D.~B., {Schinnerer}, E., {Shioya}, Y., {Shopbell}, P., {Tanaguchi}, Y.,
  {Thompson}, D., \& {Willott}, C.~J. 2009, ArXiv:0910.0444

\bibitem[{{Cardelli} {et~al.}(1989){Cardelli}, {Clayton}, \& {Mathis}}]{card89}
{Cardelli}, J.~A., {Clayton}, G.~C., \& {Mathis}, J.~S. 1989, \apj, 345, 245

\bibitem[Casertano et al.(2000)]{casertano00} {Casertano}, S., {de Mello}, D., {Dickinson}, M., {Ferguson}, H.~C., 
	{Fruchter}, A.~S., {Gonzalez-Lopezlira}, R.~A., {Heyer}, I.,
	{Hook}, R.~N., {Levay}, Z., {Lucas}, R.~A., {Mack}, J., 
	{Makidon}, R.~B., {Mutchler}, M., {Smith}, T.~E., {Stiavelli}, M., 
	{Wiggs}, M.~S. \& {Williams}, R.~E. 2000, \aj, 120, 2747

\bibitem[{{Castellano} {et~al.}(2010){Castellano}, {Fontana}, {Boutsia},
  {Grazian}, {Pentericci}, {Bouwens}, {Dickinson}, {Giavalisco}, {Santini},
  {Cristiani}, {Fiore}, {Gallozzi}, {Giallongo}, {Maiolino}, {Mannucci},
  {Menci}, {Moorwood}, {Nonino}, {Paris}, {Renzini}, {Rosati}, {Salimbeni},
  {Testa}, \& {Vanzella}}]{castellano10}
{Castellano}, M., {Fontana}, A., {Boutsia}, K., {Grazian}, A., {Pentericci},
  L., {Bouwens}, R., {Dickinson}, M., {Giavalisco}, M., {Santini}, P.,
  {Cristiani}, S., {Fiore}, F., {Gallozzi}, S., {Giallongo}, E., {Maiolino},
  R., {Mannucci}, F., {Menci}, N., {Moorwood}, A., {Nonino}, M., {Paris}, D.,
  {Renzini}, A., {Rosati}, P., {Salimbeni}, S., {Testa}, V., \& {Vanzella}, E.
  2010, \aap, 511, A20+

\bibitem[Chary(2008)]{chary08} Chary R.-R. 2008, \apj, 680, 32

\bibitem[Dressel et al.(2010)]{dressel} Dressel, L., Wong, M.H., Pavlovsky, C., and Long, K. et al., 2010. “Wide Field Camera 3 Instrument Handbook, Version 2.1” (Baltimore: STScI)

\bibitem[{{Finkelstein} {et~al.}(2010){Finkelstein}, {Papovich}, {Giavalisco},
  {Reddy}, {Ferguson}, {Koekemoer}, \& {Dickinson}}]{finkelstein10}
{Finkelstein}, S.~L., {Papovich}, C., {Giavalisco}, M., {Reddy}, N.~A.,
  {Ferguson}, H.~C., {Koekemoer}, A.~M., \& {Dickinson}, M. 2010, \apj, 719,
  1250

\bibitem[Henry et al.(2009)]{henry09} {Henry}, A.~L., {Siana},
    B., {Malkan}, M.~A., {Ashby}, M.~L.~N.,  {Bridge}, C.~R., {Chary}, {R.-R.}, {Colbert}, J.~W., 
	{Giavalisco}, M., {Teplitz}, H.~I., {McCarthy}, P.~J. 2009,
        \apj, 697, 1128

\bibitem[{{Koekemoer} {et~al.}(2002){Koekemoer}, {Fruchter}, {Hook}, \&
  {Hack}}]{drizzle}
{Koekemoer}, A.~M., {Fruchter}, A.~S., {Hook}, R.~N., \& {Hack}, W. 2002, in
  The 2002 HST Calibration Workshop : Hubble after the Installation of the ACS
  and the NICMOS Cooling System, ed. {S.~Arribas, A.~Koekemoer, \&
  B.~Whitmore}, 337--+

\bibitem[{{Komatsu} {et~al.}(2010){Komatsu}, {Smith}, {Dunkley}, {Bennett},
  {Gold}, {Hinshaw}, {Jarosik}, {Larson}, {Nolta}, {Page}, {Spergel},
  {Halpern}, {Hill}, {Kogut}, {Limon}, {Meyer}, {Odegard}, {Tucker}, {Weiland},
  {Wollack}, \& {Wright}}]{wmap7}
{Komatsu}, E., {Smith}, K.~M., {Dunkley}, J., {Bennett}, C.~L., {Gold}, B.,
  {Hinshaw}, G., {Jarosik}, N., {Larson}, D., {Nolta}, M.~R., {Page}, L.,
  {Spergel}, D.~N., {Halpern}, M., {Hill}, R.~S., {Kogut}, A., {Limon}, M.,
  {Meyer}, S.~S., {Odegard}, N., {Tucker}, G.~S., {Weiland}, J.~L., {Wollack},
  E., \& {Wright}, E.~L. 2010, ArXiv:1001.4538

\bibitem[Knapp et al.(2004)]{knapp04} Knapp, G.~R. et al. 2004, \aj, 127, 3553

\bibitem[{{Labb\'{e}} {et~al.}(2010){Labbe}, {Gonzalez}, {Bouwens}, {Illingworth},
  {Oesch}, {van Dokkum}, {Carollo}, {Franx}, {Stiavelli}, {Trenti}, {Magee}, \&
  {Kriek}}]{labbe09a}
{Labb\'{e}}, I., {Gonzalez}, V., {Bouwens}, R.~J., {Illingworth}, G.~D., {Oesch},
  P.~A., {van Dokkum}, P.~G., {Carollo}, C.~M., {Franx}, M., {Stiavelli}, M.,
  {Trenti}, M., {Magee}, D., \& {Kriek}, M. 2010, ApJL, 708, 26

\bibitem[{{Lehnert} {et~al.}(2010){Lehnert}, {Nesvadba}, {Cuby}, {Swinbank},
  {Morris}, {Cl{\'e}ment}, {Evans}, {Bremer}, \& {Basa}}]{lehnert10}
{Lehnert}, M.~D., {Nesvadba}, N.~P.~H., {Cuby}, J., {Swinbank}, A.~M.,
  {Morris}, S., {Cl{\'e}ment}, B., {Evans}, C.~J., {Bremer}, M.~N., \& {Basa},
  S. 2010, \nat, 467, 940

\bibitem[{{Malhotra} {et~al.}(2005){Malhotra}, {Rhoads}, {Pirzkal}, {Haiman},
  {Xu}, {Daddi}, {Yan}, {Bergeron}, {Wang}, {Ferguson}, {Gronwall},
  {Koekemoer}, {Kuemmel}, {Moustakas}, {Panagia}, {Pasquali}, {Stiavelli},
  {Walsh}, {Windhorst}, \& {di Serego Alighieri}}]{malhotra05}
{Malhotra}, S., {Rhoads}, J.~E., {Pirzkal}, N., {Haiman}, Z., {Xu}, C.,
  {Daddi}, E., {Yan}, H., {Bergeron}, L.~E., {Wang}, J., {Ferguson}, H.~C.,
  {Gronwall}, C., {Koekemoer}, A., {Kuemmel}, M., {Moustakas}, L.~A.,
  {Panagia}, N., {Pasquali}, A., {Stiavelli}, M., {Walsh}, J., {Windhorst},
  R.~A., \& {di Serego Alighieri}, S. 2005, \apj, 626, 666

\bibitem[{{McLure} {et~al.}(2010){McLure}, {Dunlop}, {Cirasuolo}, {Koekemoer},
  {Sabbi}, {Stark}, {Targett}, \& {Ellis}}]{mclure09}
{McLure}, R.~J., {Dunlop}, J.~S., {Cirasuolo}, M., {Koekemoer}, A.~M., {Sabbi},
  E., {Stark}, D.~P., {Targett}, T.~A., \& {Ellis}, R.~S. 2010, MNRAS, 403, 960

\bibitem[{{Oesch} {et~al.}(2010{\natexlab{a}}){Oesch}, {Bouwens}, {Carollo},
  {Illingworth}, {Trenti}, {Stiavelli}, {Magee}, {Labb{\'e}}, \&
  {Franx}}]{oesch09_size}
{Oesch}, P.~A., {Bouwens}, R.~J., {Carollo}, C.~M., {Illingworth}, G.~D.,
  {Trenti}, M., {Stiavelli}, M., {Magee}, D., {Labb{\'e}}, I., \& {Franx}, M.
  2010{\natexlab{a}}, \apjl, 709, L21

\bibitem[{{Oesch} {et~al.}(2010{\natexlab{b}}){Oesch}, {Bouwens},
  {Illingworth}, {Carollo}, {Franx}, {Labb{\'e}}, {Magee}, {Stiavelli},
  {Trenti}, \& {van Dokkum}}]{oesch09_zdrop}
{Oesch}, P.~A., {Bouwens}, R.~J., {Illingworth}, G.~D., {Carollo}, C.~M.,
  {Franx}, M., {Labb{\'e}}, I., {Magee}, D., {Stiavelli}, M., {Trenti}, M., \&
  {van Dokkum}, P.~G. 2010{\natexlab{b}}, \apjl, 709, L16

\bibitem[{{Oesch} {et~al.}(2007){Oesch}, {Stiavelli}, {Carollo}, {Bergeron},
  {Koekemoer}, {Lucas}, {Pavlovsky}, {Trenti}, {Lilly}, {Beckwith}, {Dahlen},
  {Ferguson}, {Gardner}, {Lacey}, {Mobasher}, {Panagia}, \& {Rix}}]{oesch07}
{Oesch}, P.~A., {Stiavelli}, M., {Carollo}, C.~M., {Bergeron}, L.~E.,
  {Koekemoer}, A.~M., {Lucas}, R.~A., {Pavlovsky}, C.~M., {Trenti}, M.,
  {Lilly}, S.~J., {Beckwith}, S.~V.~W., {Dahlen}, T., {Ferguson}, H.~C.,
  {Gardner}, J.~P., {Lacey}, C., {Mobasher}, B., {Panagia}, N., \& {Rix}, H.
  2007, \apj, 671, 1212

\bibitem[Oke(1974)]{oke} Oke, J.~B. 1974, ApJS, 27, 21 

\bibitem[{{Ouchi} {et~al.}(2009){Ouchi}, {Mobasher}, {Shimasaku}, {Ferguson},
  {Fall}, {Ono}, {Kashikawa}, {Morokuma}, {Nakajima}, {Okamura}, {Dickinson},
  {Giavalisco}, \& {Ohta}}]{ouchi09}
{Ouchi}, M., {Mobasher}, B., {Shimasaku}, K., {Ferguson}, H.~C., {Fall}, S.~M.,
  {Ono}, Y., {Kashikawa}, N., {Morokuma}, T., {Nakajima}, K., {Okamura}, S.,
  {Dickinson}, M., {Giavalisco}, M., \& {Ohta}, K. 2009, \apj, 706, 1136

\bibitem[{{Reid} {et~al.}(2001){Reid}, {Burgasser}, {Cruz}, {Kirkpatrick}, \&
  {Gizis}}]{reid01}
{Reid}, I.~N., {Burgasser}, A.~J., {Cruz}, K.~L., {Kirkpatrick}, J.~D., \&
  {Gizis}, J.~E. 2001, \aj, 121, 1710

\bibitem[Robertson et al.(2010)]{robertson10} Robertson, B.~E.,
  Ellis, R.~S., Dunlop, J.~S., McLure, R.~J. \& Stark, D.~P., 2010, \nat, 468, 49

\bibitem[{{Schlegel} {et~al.}(1998){Schlegel}, {Finkbeiner}, \&
  {Davis}}]{schlegel98}
{Schlegel}, D.~J., {Finkbeiner}, D.~P., \& {Davis}, M. 1998, \apj, 500, 525

\bibitem[{{Steidel} {et~al.}(1996){Steidel}, {Giavalisco}, {Pettini},
  {Dickinson}, \& {Adelberger}}]{steidel96}
{Steidel}, C.~C., {Giavalisco}, M., {Pettini}, M., {Dickinson}, M., \&
  {Adelberger}, K.~L. 1996, \apjl, 462, L17+

\bibitem[{{Trenti} \& {Stiavelli}(2008)}]{trenti08}
{Trenti}, M. \& {Stiavelli}, M. 2008, \apj, 676, 767

\bibitem[{{Trenti} {et~al.}(2010){Trenti}, {Stiavelli}, {Bouwens}, {Oesch},
  {Shull}, {Illingworth}, {Bradley}, \& {Carollo}}]{trenti10b}
{Trenti}, M., {Stiavelli}, M., {Bouwens}, R.~J., {Oesch}, P., {Shull}, J.~M.,
  {Illingworth}, G.~D., {Bradley}, L.~D., \& {Carollo}, C.~M. 2010, \apjl, 714,
  L202

\bibitem[Wilkins et al.(2010)]{wilkins10} 
	Wilkins, S.~M., Bunker, A.~J., Ellis, R.~S., Stark, D.,
        Stanway, E.~R.; Chiu, K., Lorenzoni, S. \& Jarvis, M.~J. 2010,
        \mnras, 403, 938

\bibitem[{{Yan} {et~al.}(2010){Yan}, {Yan}, {Zamojski}, {Windhorst},
  {McCarthy}, {Fan}, {R{\"o}ttgering}, {Koekemoer}, {Robertson}, {Dav{\'e}}, \&
  {Cai}}]{yan10}
{Yan}, H., {Yan}, L., {Zamojski}, M.~A., {Windhorst}, R.~A., {McCarthy}, P.~J.,
  {Fan}, X., {R{\"o}ttgering}, H.~J.~A., {Koekemoer}, A.~M., {Robertson},
  B.~E., {Dav{\'e}}, R., \& {Cai}, Z. 2010, ArXiv:1010:2261

\end{thebibliography}

\clearpage
  

\newpage 
\begin{deluxetable}{lrrrrrrrrrrrr}
\small
\tablecolumns{13} 
\tablecaption{BoRG fields location, exposure times and $5\sigma$ magnitude limits ($r=0.32\mathrm{"}$)\label{tab:fields}}
\tablehead{ \colhead{{\small{Field}}} & \colhead{\small RA}  &
  \colhead{\small{DEC}} &\multicolumn{2}{c}{F125W}
  &\multicolumn{2}{c}{F160W} & \multicolumn{2}{c}{F098M} &
  \multicolumn{2}{c}{F606W} &\multicolumn{2}{c}{F600LP} \\ 
\colhead{} & \colhead {[deg]} & \colhead {[deg]} & \colhead {$t$ [s]}
&\colhead {$m_{lim}$} & \colhead {$t$ [s]} &\colhead {$m_{lim}$} &
\colhead {$t$ [s]} &\colhead {$m_{lim}$} & \colhead {$t$ [s]}
&\colhead {$m_{lim}$} & \colhead {$t$ [s]} &\colhead {$m_{lim}$} }

\small
\startdata 
   
   \small   BoRG93 &$     99.286 $&$    -75.307 $&$  2412 $&$    26.6 $&$  1612 $&$    26.0 $&$  6218 $&$    26.8 $&$  4290 $&$    26.9 $& \\
 \small        BoRG81 &$     88.277 $&$    -64.091 $&$  2612 $&$    26.9 $&$  2012 $&$    26.3 $&$  6418 $&$    27.0 $&$  3624 $&$    27.1 $& \\
 \small        BoRG73 &$    136.403 $&$      2.925 $&$  2709 $&$    27.1 $&$  1906 $&$    26.6 $&$  5518 $&$    27.0 $&$  3106 $&$    27.0 $& \\
 \small        BoRG70 &$    157.712 $&$     38.059 $&$  1506 $&$    26.3 $&$  1306 $&$    26.1 $&$  3109 $&$    26.4 $&$  1815 $&$    26.4 $& \\
   \small      BoRG66 &$    137.284 $&$     -0.030 $&$  1806 $&$    26.8 $&$  1006 $&$    26.1 $&$  3909 $&$    26.9 $&$  2650 $&$    26.9 $& \\
   \small      BoRG58 &$    219.230 $&$     50.719 $&$  2509 $&$    27.0 $&$  1806 $&$    26.6 $&$  4912 $&$    27.1 $&$  2754 $&$    26.8 $& \\
 \small        BoRG49 &$    191.184 $&$     33.937 $&$  1506 $&$    26.6 $&$  1106 $&$    26.2 $&$  3409 $&$    26.8 $&$  1789 $&$    26.8 $& \\
  \small       BoRG45 &$    141.390 $&$     40.005 $&$  1106 $&$    26.1 $&$   903 $&$    25.9 $&$  2806 $&$    26.2 $&$  1276 $&$    26.1 $& \\
 \small        BoRG39 &$    138.567 $&$     28.363 $&$  2206 $&$    26.9 $&$  1706 $&$    26.5 $&$  4615 $&$    26.9 $&$  2571 $&$    26.9 $& \\
  \small       BoRG30\tablenotemark{!} &$    125.011 $&$     23.536 $&$   703 $&$    26.1 $&$   703 $&$    25.8 $&$  3109 $&$    26.5 $&$  2556 $&$    26.8 $& \\
  \small       BoRG2t\tablenotemark{!} &$     95.903 $&$    -64.528 $&$  1206 $&$    26.4 $&$   503 $&$    25.6 $&$  1806 $&$    26.4 $&$  2133 $&$    26.7 $& \\
 \small        BoRG2n &$     84.879 $&$    -64.153 $&$  2309 $&$    26.6 $&$  1406 $&$    26.3 $&$  4112 $&$    26.6 $&$  3171 $&$    27.1 $& \\
  \small       BoRG2k &$     95.952 $&$    -64.665 $&$  1206 $&$    26.6 $&$   906 $&$    26.0 $&$  2909 $&$    26.7 $&$  2135 $&$    26.9 $& \\
  \small       BoRG1v &$    187.473 $&$      7.825 $&$  1806 $&$    25.9 $&$  1406 $&$    25.6 $&$  4112 $&$    26.0 $&$  2436 $&$    26.6 $& \\
  \small       BoRG1r &$    140.403 $&$     45.087 $&$  2106 $&$    26.6 $&$  1706 $&$    26.3 $&$  4812 $&$    26.7 $&$  2708 $&$    26.6 $& \\
 \small        BoRG1n &$    122.086 $&$     39.759 $&$  2206 $&$    26.5 $&$  1406 $&$    26.0 $&$  4612 $&$    26.5 $&$  2600 $&$    26.7 $& \\
 \small        BoRG1k &$    247.889 $&$     37.610 $&$  1206 $&$    26.6 $&$   906 $&$    26.1 $&$  2909 $&$    26.8 $&$  1260 $&$    26.3 $& \\
 \small        BoRG0y &$    177.963 $&$     54.684 $&$  2809 $&$    27.1 $&$  1906 $&$    26.7 $&$  6021 $&$    27.1 $&$  2898 $&$    27.1 $& \\
 \small        BoRG0j &$    178.180 $&$      0.933 $&$  2209 $&$    26.7 $&$  1606 $&$    26.5 $&$  4515 $&$    26.8 $&$  2647 $&$    26.8 $& \\
  \small       BoRG0c &$    118.986 $&$     30.718 $&$  1906 $&$    26.6 $&$  1406 $&$    26.3 $&$  4712 $&$    26.8 $&$  2600 $&$    26.9 $& \\
  \small       BoRG0g &$    124.834 $&$     49.183 $&$  1206 $&$    26.5 $&$   806 $&$    25.8 $&$  3009 $&$    26.6 $&$  1908 $&$    26.5 $& \\
 \small   BoRG0p&$    182.358 $&$     45.723 $&$  3709 $&$    27.3 $&$  2909 $&$
27.0 $&$ 13729 $&$    27.6 $&$  2707 $&$    27.0 $& $2234$& $26.4$\\
 \small   BoRG0t &$    117.707 $&$     29.282 $&$  5115 $&$    27.2 $&$3912 $&$
26.9 $&$ 18641 $&$    27.5 $&$  2826 $&$    26.9 $& $3752$ & $26.7$\\
  \small    yan11 &$    190.553 $&$     57.268 $&$  2509 $&$    26.9 $&$  2309 $&$    26.6 $&$  5215 $&$    27.0 $& &&$  2800 $&$    26.5 $ \\
 \small         yan19 &$    204.200 $&$     -0.462 $&$  1203 $&$    26.7 $&$  1203 $&$    26.3 $&$  6818 $&$    27.1 $& &&$  2270 $&$    26.2 $ \\
   \small       yan24 &$     33.408 $&$     12.914 $&$  1403 $&$    26.0 $&$  1403 $&$    25.8 $&$  2806 $&$    26.0 $& &&$  2294 $&$    26.0 $ \\
 \small         yan28 &$    141.379 $&$     44.427 $&$  1603 $&$    26.8 $&$  1403 $&$    26.5 $&$  6012 $&$    27.0 $& &&$  2374 $&$    26.4 $ \\
   \small       yan32 &$    205.128 $&$     41.386 $&$  3206 $&$    27.2 $&$  2806 $&$    27.0 $&$ 17435 $&$    27.6 $& &&$  3810 $&$    26.8 $ \\
   \small       yan51 &$    231.038 $&$      9.905 $&$  1603 $&$    26.6 $&$       1303 $&$    26.2 $&$  8718 $&$    27.0 $& &&$  2078 $&$    26.3       $ \\

\enddata \tablecomments{Survey area: approximately
  $130~\mathrm{arcmin^2}$. Effective area for F098M-dropout detection:
  approximately $97~\mathrm{arcmin^2}$.}
\tablenotetext{!}{Data missing due to scheduling constraint/conflict.}
\end{deluxetable}

\newpage

\begin{deluxetable}{lrrrrrrrrrr}
\small
\tablecolumns{11} 
\tablecaption{BoRGs F098M-dropouts \label{tab:candidates}}
\tablehead{ \colhead{~} & & & \colhead{\small $m_{J}$}  & \multicolumn{2}{l}{ {\small IR-Colors} } &  \multicolumn{4}{l}{ {\small{ S/N}}\tablenotemark{a}} & \colhead{\small{Stellarity}} \\
\colhead{} &\colhead{\small RA}  &\colhead{DEC}  & \colhead{} & \colhead{Y-J} & \colhead{J-H} & \colhead{V} & \colhead{Y} & \colhead{J} & \colhead{H} & \colhead{} } 
\startdata 
{\tiny BoRG66\_1741-1157}  &  $137.2732$ &  $-0.0297$ & $26.2 \pm 0.2$ & $1.9\pm 0.6$ & $-0.4 \pm 0.4$ & -0.8 & 2.1 & 8.7 & 3.3 & 0.01 \\
{\tiny BoRG1k\_0847-0733}\tablenotemark{+} &  $247.8968$ & $+37.6039$ &$25.5 \pm 0.1$ & $1.9 \pm 0.4$ & $0.0 \pm 0.2$ & 1.3 & 2.4 & 11.4 & 7.4 & 0.03 \\
{\tiny BoRG0t\_0958-0641}\tablenotemark{+}\tablenotemark{b}& $117.7142$  & $+29.2715$ & $26.7 \pm 0.2$ & $\geq 2.6$ & $0.0^{+0.2}_{-0.5}$ & -0.9 & 0.1 & 8.6 & 6.8 & 0.79 \\
{\tiny BoRG58\_1787-1420}\tablenotemark{+} & 219.2107 & +50.7260 & $25.8 \pm 0.1$& $\geq 2.8$ & $-0.1\pm 0.2$ & -1.4 & -0.7 & 13.2 & 8.0 & 0.36 \\
\enddata \tablecomments{Photometric properties of candidates and their
  coordinates (Deg, J2000 system).  Magnitude $m_J$ is AUTOMAG from
  SExtractor. Colors are derived from ISOMAG (fluxes below $1\sigma$
  have been replaced with $1\sigma$ limit). SExtractor Stellarity
  parameter also reported.} \tablenotetext{+}{In \citet{yan10}
  catalog.}  \tablenotetext{a}{Negative $S/N$ is due to negative
  sky-subtracted flux within the photometric aperture used.}
\tablenotetext{b}{Also available: F600LP filter with $S/N =
  -0.2$.} \end{deluxetable}


\end{document}